\documentclass[twocolumn]{aastex63}
\usepackage{xcolor}
\usepackage{hyperref}
\usepackage{amsmath,amssymb,mathrsfs}
\usepackage{soul}
\usepackage{booktabs}

\def \psr {\mbox{1RXS\, J141256.0$+$792204}}
\def \xmm {XMM-Newton}

\def \nic {NICER}

\def \cha {Chandra}

\def \pdot {\dot P}

\def\nudot {\dot \nu}
\def\nuddot {\ddot \nu}
\def\brake {\nu\ddot\nu \dot\nu^{-2}}

\def\ltsima{$\; \buildrel < \over \sim \;$}
\def\lsim{\lower.5ex\hbox{\ltsima}}
\def\gtsima{$\; \buildrel > \over \sim \;$}
\def\gsim{\lower.5ex\hbox{\gtsima}}

\graphicspath{{fig/}}

\defcitealias{Davies81}{DP81} 

\received{2024 August 2}
\revised{2024 October 14}
\accepted{2024 October 28}
\submitjournal{ApJ}

\shorttitle{Calvera Proper Motion}
\shortauthors{Rigoselli et al.}

\begin{document}
\title{The Proper Motion of the High Galactic Latitude Pulsar Calvera}

\correspondingauthor{Michela Rigoselli}
\email{michela.rigoselli@inaf.it}

\author[0000-0001-6641-5450]{M.~Rigoselli} 
\affiliation{INAF -- Osservatorio Astronomico di Brera,
Via Brera 28, I-20121 Milano, Italy}
\affiliation{INAF -- Istituto di Astrofisica Spaziale e Fisica Cosmica, Via A. Corti 12, I-20133 Milano, Italy}

\author[0000-0003-3259-7801]{S.~Mereghetti} 
\affiliation{INAF -- Istituto di Astrofisica Spaziale e Fisica Cosmica, Via A. Corti 12, I-20133 Milano, Italy}

\author[0000-0003-4814-2377]{J.~P. Halpern}
\affiliation{Columbia Astrophysics Laboratory, Columbia University, 550 West 120th Street, New York, NY 10027, USA}

\author[0000-0003-3847-3957]{E.~V. Gotthelf}
\affiliation{Columbia Astrophysics Laboratory, Columbia University, 550 West 120th Street, New York, NY 10027, USA}

\author[0000-0002-1429-9010]{C.~G. Bassa}
\affiliation{ASTRON Netherlands Institute for Radio Astronomy, Oude Hoogeveensedijk 4, 7991 PD Dwingeloo, The Netherlands}

\begin{abstract}
Calvera (\psr) is a  pulsar of characteristic age 285~kyr at a high Galactic latitude of $b=+37^{\circ}$, detected only in soft thermal X-rays.
We measure a new and precise proper motion for Calvera using \cha\ HRC-I observations obtained 10 years apart.
We also derive a new phase-connected ephemeris using 6 years of NICER data, including the astrometric position and proper motion as fixed parameters in the timing solution.
Calvera is located near the center of a faint, circular radio ring that was recently discovered by LOFAR and confirmed as a supernova remnant (SNR) by the detection of $\gamma$-ray emission with Fermi/LAT.
The proper motion of $78.5 \pm 2.9$~mas\,yr$^{-1}$ at position angle $241^{\circ}\!\! .3 \pm 2^{\circ}\!\! .2$ (in Galactic coordinates) points away from the center of the ring, a result which differs markedly from a previous low-significance measurement, and greatly simplifies the interpretation of the SNR/pulsar association.
It argues that the supernova indeed birthed Calvera $<10$~kyr ago, with an initial spin period close to its present value of 59~ms.
The tangential velocity of the pulsar depends on its uncertain distance, $v_t=(372 \pm 14)~d_{1\,{\rm kpc}}$ km\,s$^{-1}$, but is probably dominated by the supernova kick, while its progenitor could have been a runaway O or B star from the Galactic disk.

\medskip\noindent
{\it Unified Astronomy Thesaurus concepts:} Pulsars (1306); Neutron stars (1108); Proper motions (1295)

\end{abstract}

\medskip\noindent


\section{Introduction}
\label{sec:intro}

\begin{figure}
\centering
\includegraphics[width=1\columnwidth]{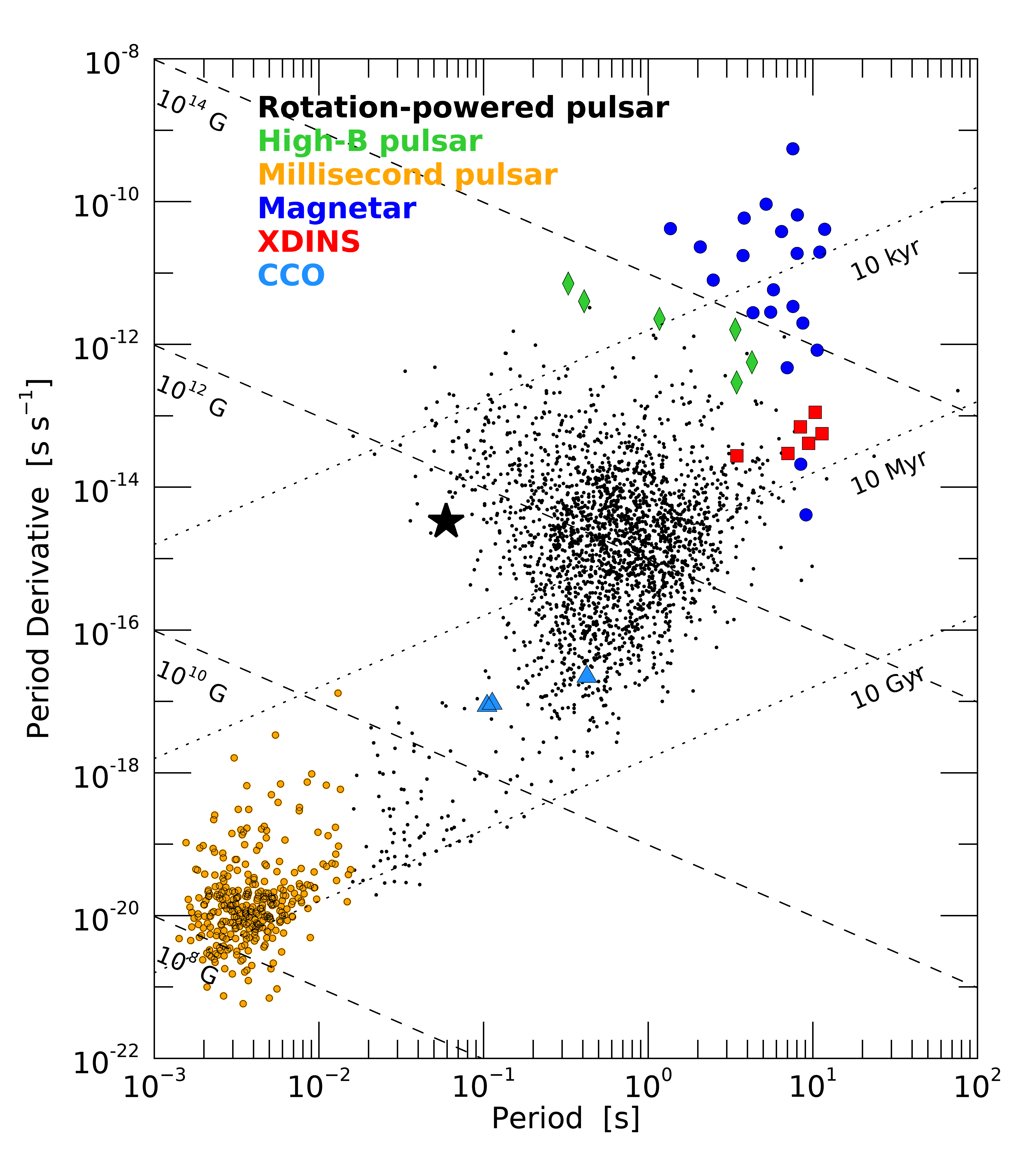}

  \caption{$P$-$\pdot$ diagram of pulsars. The bulk of the pulsar population is made up of the rotation-powered pulsars (black dots), sub-divided into high-B pulsars (green diamonds) and millisecond pulsars (orange dots). The other classes of isolated neutron stars plotted here are the magnetars (blue dots), the XDINSs (red squares) and the CCOs (light blue triangles). The position of Calvera is indicated by the black star.
  Lines of equal characteristic age (dotted, $10^{4} - 10^{10}$ yr) and equal dipole magnetic field (dashed, $10^{8} - 10^{14}$ G) are indicated.
  The data are taken from the ATNF Pulsar Catalogue, version 2.5.1 \citep{man05}.
   \label{fig:ppdot}
 }
\end{figure}

\psr\ is a pulsar with a unique combination of properties that resist pigeonholing into any established class of isolated neutron star (INS).
It was discovered in the ROSAT All Sky Survey as a soft X-ray source with high X-ray-to-optical flux ratio, clearly pointing to an INS nature \citep{rut08}.
Its very soft thermal X-ray spectrum (blackbody temperature $\approx\!0.2$~keV) and lack of radio emission \citep{hes07} suggested membership in the small class of thermally emitting, radio-quiet INS known as X-ray dim isolated neutron stars (XDINS, see \citealt{tur09} for a review), also called ``Magnificent Seven'', hence the nickname ``Calvera''.
However, this hypothesis was discarded after the  discovery of its short spin period of $P=59$~ms \citep{zan11} and period derivative $\pdot=3.2\times10^{-15}$ \citep{hal13}.
These values do not fit with those of the XDINS, which have periods in the range $\sim\!3-17$~s and $\pdot$ of a few $\times10^{-14}$. 
The timing parameters of Calvera correspond to a characteristic age $\tau_c \equiv P/2\dot P=2.85\times10^5$~yr, a spin-down power $\dot E=6.3\times10^{35}$ erg~s$^{-1}$ and, with the usual dipole spin-down assumptions, a surface dipole magnetic field strength $B_s=4.4\times10^{11}$~G, close to those of ordinary rotation-powered pulsars rather than the high $B$-field XDINSs (see Figure~\ref{fig:ppdot}). 

The X-ray luminosity of Calvera, $L_X\approx1.3\times10^{32}~d_{1\,\rm kpc}^2$ erg~s$^{-1}$, is poorly constrained due to its unknown distance, which, in the absence of detection at other wavelengths, can only be estimated by modeling its thermal X-ray emission \citep{she09,zan11,shi16,bog19}. 
The latest and most detailed analysis showed that both the phase-resolved X-ray spectra and the energy dependence of the pulsed fraction are well reproduced by a hydrogen atmosphere model with the latitudinal temperature variation expected for a dipolar field, plus a small, hotter polar cap \citep{mer21}. 
The  best fit of this model gives a distance of $\approx\!3.3$~kpc, which, considering the source's Galactic coordinates ($\ell,b)=(118^{\circ}\!\! .32,+37^{\circ}\!\! .02$), would correspond to a height of $\approx\!2$~kpc above the Galactic plane.
If the neutron star was born from a young, massive star in the Galactic disk and its true age is $\sim\!\tau_c$, this would imply a projected velocity of $v_t\sim6800$~km\,s$^{-1}$, unreasonably large to be the result of a supernova kick. 

A new development that may reveal the origin of Calvera is the discovery with LOFAR of a diffuse ring of faint radio (144 MHz) emission surrounding the pulsar \citep{2022A&A...667A..71A}. 
The ring has a diameter of $\approx\!1^{\circ}$, while Calvera is $\approx\!5^{\prime}$ from its center.
Its properties are consistent with a supernova remnant (SNR) expanding in the low-density interstellar medium expected at this location far from the Galactic plane.
This is also supported by the detection with Fermi-LAT of extended $\gamma$-ray emission with spectral properties consistent with those of a young SNR 
\citep{2022ApJ...941..194X,2023MNRAS.518.4132A}.

Given the rarity of both pulsars and SNRs at this Galactic latitude, the probability is high that the two objects are associated \citep{2022A&A...667A..71A}. 
However, a proper motion measurement of Calvera using Chandra \citep{hal15} did not point back to the center of the ring, and the significance of the proper motion was marginal, leaving the association inconclusive.

In order to obtain a much more precise proper motion for Calvera, we performed a new \cha\ observation to be compared with the similar image obtained 10 years earlier. The proper motion measurement is described in Section~\ref{sec:pm}, while in Section~\ref{sec:timing} we report timing analysis of NICER observations taking into account the new proper motion.
The implications of our new results for the origin of this unusual pulsar are discussed in Section~\ref{sec:discussion}.

\setlength{\tabcolsep}{0.6em}
\begin{table*}
\centering \caption{Log of \cha\ HRC-I Observations and Background-subtracted Counts}
\label{tab:log}

\begin{tabular}{rccccc}
\toprule
\midrule
ObsID & Start time & End time  & Exposure & AGN & Calvera \\[3pt]
  &  (UTC)     &  (UTC)    & (ks) & (Net Counts)
  &  (Net Counts) \\[3pt]
\midrule
8508 & 2007-02-18 17:49:14 & 2007-02-18 18:51:51 & 2.14 & $8\pm3$ & $189\pm14$ \\
15806 & 2014-04-02 05:46:35 & 2014-04-02 14:37:46 & 29.96 & $174\pm14$ & $2783\pm54$\\
26623 & 2024-03-04 02:02:16 & 2024-03-04 06:48:08 & 14.18 & $96\pm10$ & $1109\pm34$ \\
29305 & 2024-03-05 18:30:34 & 2024-03-05 23:16:12 & 14.36 & $89\pm10$ & $1124\pm34$\\

\bottomrule\\[-5pt]
\end{tabular}

\raggedright
\end{table*}

\section{Data analysis and results} 

\subsection{Proper motion}
\label{sec:pm}

Proper motion measurements in X-rays are best done with the Chandra High Resolution Camera (HRC-I), which collects $0.1-10$~keV photons with little or no energy resolution that are binned into pixels of $0^{\prime\prime}\!\! .1318$ over a $30^{\prime}\times30^{\prime}$ field of view.  Its spatial resolution is $\approx\!0.\!^{\prime\prime}4$ on axis. Optically identified X-ray sources in the image are typically used to correct the astrometry to greater accuracy than that provided by the on-board star trackers. 

Calvera itself is by far the strongest source in its image, and there is only one reference source that is bright enough and close enough to Calvera to provide a useful correction.  It is an AGN identified as USNO-B1.0 1693$-$0051234 \citep{hal15}, which is $2^{\prime}$ south of Calvera, at R.A.=$14^{\rm h}12^{\rm m}59^{\rm s}\!\! .473644(56)$, Decl.=$+79^{\circ}19^{\prime}58^{\prime\prime}\!\! .58616(16)$ in the Gaia DR3 catalog \citep{2023A&A...674A...1G}. 
The Gaia-CRF3 celestial reference frame is equivalent to J2000.0 for our purposes.
All other sources have too few photons, or are located off axis where the point spread function (PSF) is significantly degraded. With only one reference source, translations of the reference frame can be corrected, but not rotation. However, as discussed in \citet{hal15}, the uncertainty in the roll angle of Chandra is sufficiently small that rotation does not contribute a significant error in position over the $2^{\prime}$ separation of the two sources.  

\cite{hal15} attempted to measure proper motion between images taken in 2007 and 2014 (see \dataset[DOI: 10.25574/cdc.307]{https://doi.org/10.25574/cdc.307} and Table~\ref{tab:log} for a log of all HRC-I observations of Calvera).
The precision was limited by the brevity of the 2007 observation (2~ks), in which less than ten photons were detected from the AGN.  In addition, the 2007 pointing had Calvera on axis, with the effect that the PSF of the AGN was broadened at its location $2^{\prime}$ off axis.  Thus, the precision of the proper motion measurement was limited by the statistical error on the 2007 X-ray position of the AGN. The longer observation of 2014, as well as our new one in 2024, were instead pointed at the AGN, which optimizes the PSF of the weaker source while the much brighter Calvera remains well localized. This enables a much more accurate and precise position measurement.

The new HRC-I observation consisted of two pointings with net exposure times of 14~ks each, carried out on 2024 March 4 and 5. This split was required for thermal management of the spacecraft. Thus, the total number of counts collected from the two sources in 2024 were similar to those of the single 30~ks observation in 2014.
We treated the two 2024 pointings independently, and excluded the poor 2007 observation from the following analysis.

We reprocessed the data with the tool \texttt{chandra\_repro} of the \cha\ Interactive Analysis of Observation software (\texttt{CIAO}; \citealt{2006SPIE.6270E..1VF}), version 4.16, using the calibration database \texttt{CALDB} 4.11.1.
Instead of deriving the positions of the sources with a simple centroiding method, we measured them
with a maximum likelihood analysis which used, for each source and data set, the corresponding PSF. Given that the PSF depends on the source spectral shape and position in the field of view, we used the most updated version of the \texttt{CHaRT/MARX} software\footnote{\url{https://cxc.cfa.harvard.edu/ciao/threads/marx/}.} and its task \texttt{simulate\_psf} to compute the most suitable PSFs. We adopted the spectra obtained by fitting the \xmm\ data: a multi-temperature thermal spectrum for Calvera \citep[see Table 2 of][]{mer21} and an absorbed power law with photon index $\Gamma=1.8$ and column density $N_{\rm H} = 2.5\times 10^{20}$ cm$^{-2}$ for the AGN. 
The maximum likelihood analysis was based on the method described in \citet{2018A&A...615A..73R}. In particular, the analysis was done on a region $60\times60$ pixels centered at the source position to derive the most likely values and corresponding uncertainties for the local background level, the source net counts (see Table~\ref{tab:log}) and position in detector coordinates. The latter were then converted into equatorial coordinates with \texttt{dmcoords}.

The X-ray positions measured in this way for the two sources are listed in Table~\ref{tab:pos}, together with the position of Calvera corrected for the difference between the Gaia position of the AGN and its X-ray position. The errors on X-ray positions are computed in a standard way by the maximum likelihood algorithm \citep[see, e.g.,][]{1996ApJ...461..396M}, and are by far the dominant source of uncertainty, since they are $\sim\!100$ times larger than those on the Gaia position. These uncertainties are then propagated when differences on positions are computed.
The corrected positions of Calvera require a significant proper motion as listed in the last columns of  Table~\ref{tab:pos}.  Then, by averaging the results of the two 2024 pointings, we obtain the final proper motion components,
$\mu_{\alpha}{\rm cos}\,\delta=78.1\pm2.9$ mas\,yr$^{-1}$ and $\mu_{\delta}=8.0\pm3.0$ mas\,yr$^{-1}$. The total proper motion is $\mu=78.5\pm2.9$ mas\,yr$^{-1}$ at position angle $84^{\circ}\!\! .2\pm2^{\circ}\!\! .2$ east of north, or $241^{\circ}\!\! .3$ in Galactic coordinates.

\setlength{\tabcolsep}{0.2em}
\begin{table*}
\centering \caption{HRC-I Position Measurements}
\label{tab:pos}

\footnotesize
\begin{tabular}{ccccccc}
\toprule
\midrule
ObsID & Epoch & AGN (measured) & Calvera (measured) & Calvera (corrected) & $\mu_{\alpha}\cos\delta$  & $\mu_{\delta}$   \\[3pt]
 & (year)  & R.A. (h m s) \ \ \ \ Decl. ($^\circ\ ^{\prime}\ ^{\prime\prime}$) & R.A. (h m s) \ \ \ \ Decl. ($^\circ\ ^{\prime}\ ^{\prime\prime}$)  &   R.A. (h m s) \ \ \ \ Decl. ($^\circ\ ^{\prime}\ ^{\prime\prime}$) & (mas\,yr$^{-1}$) & (mas\,yr$^{-1}$)   \\[3pt]
\midrule
15806 & 2014.250   & 14 12 59.464(9)  +79 19 58.36(2)  & 14 12 55.838(2) +79 22 03.240(8)  & 14 12 55.847(9)  +79 22 03.46(3) & \dots\ & \dots \\
26623 & 2024.255   & 14 12 59.109(10) +79 19 59.07(3)  & 14 12 55.755(3) +79 22 04.063(13) & 14 12 56.119(11) +79 22 03.57(3) & $76.0\pm3.9$   &  $11.1\pm4.0$  \\
29305 & 2024.259   & 14 12 59.362(12) +79 19 58.49(4)  & 14 12 56.025(4) +79 22 03.415(13) & 14 12 56.137(13) +79 22 03.50(4) & $80.8\pm4.4$   &  $4.0\pm4.6$  \\

\bottomrule\\[-5pt]
\end{tabular}
\raggedright
\end{table*}

\subsection{Timing analysis}
\label{sec:timing}

We analyzed all the data of Calvera available in the public archive of   \nic\ observations. They span  rather uniformly the time interval from 2017 September 15 to 2023 November 2.
We reduced the data with the \texttt{nicerdas} software (version 10a), including all of the most recently released patches and calibration files (\texttt{CALDB XTI20221001}).  
We filtered the data using the tool  \texttt{nicerl2} and applied the standard cuts. 
To exclude time intervals of high particle background, we selected only the observations with  $K_p < 5$ and \texttt{COR\_SAX} $> 1.914~K_p^{0.684}+0.25$, where $K_p$ is an indicator of the effect of solar wind activity on the Earth's magnetosphere and the condition on \texttt{COR\_SAX} excludes parts of the orbit in regions with low cut-off rigidity. 
These cuts resulted in 682 observations for a total exposure time of 1.5 Ms spanning 6.14 yr.

The photon arrival times were converted to barycentric dynamical time (TDB) using the DE430 JPL ephemeris, the TT(BIPM2021) clock realization, and the time-variable pulsar position resulting from the proper motion measured in Section~\ref{sec:pm} and listed in Table~\ref{tab:timing}.

We grouped the observations in order to have at least 5 ks of exposure time in each group, but we restricted the time span of each group to less than 7 days. 
The pulse time of arrival (ToA) for each of the 223 resulting groups was determined by fitting a sine function to the folded pulse profile in the $0.4-2$~keV range.

\begin{figure*}
\centering
\includegraphics[width=1\textwidth]{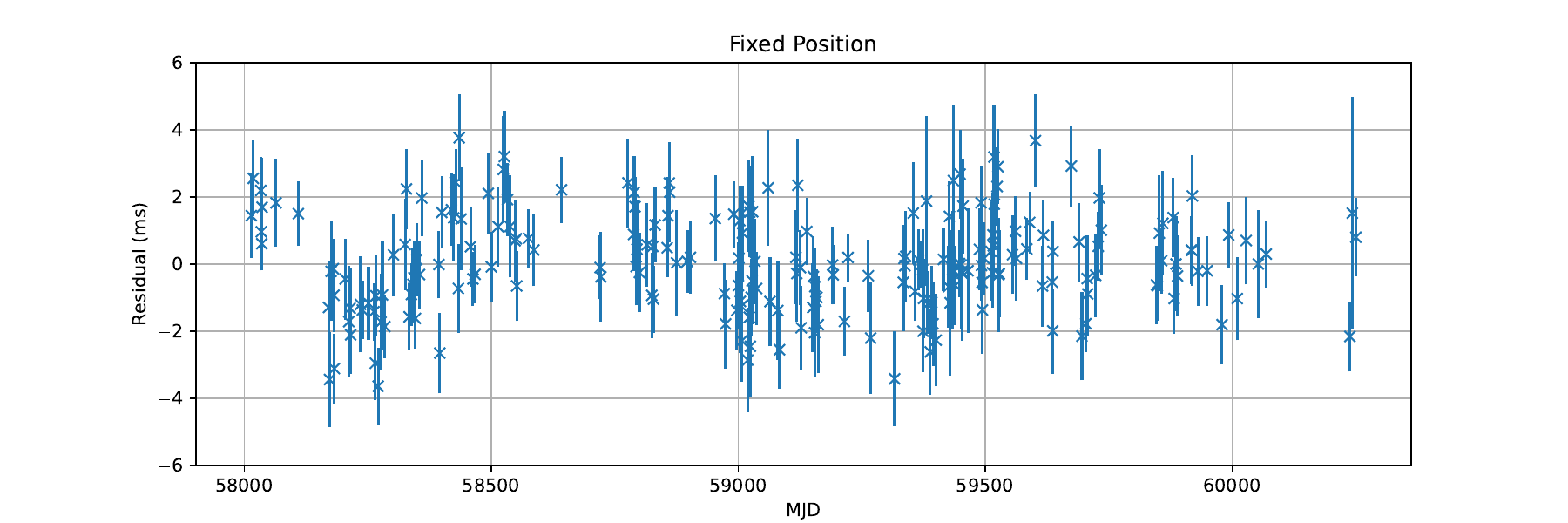}
\includegraphics[width=1\textwidth]{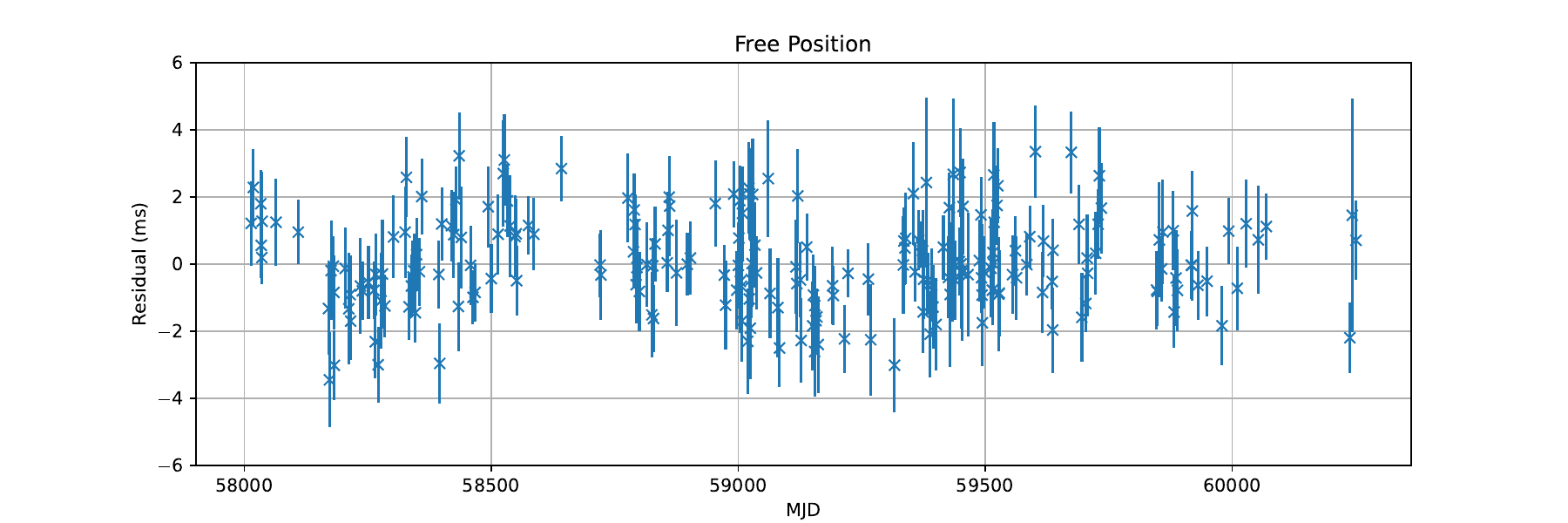}

  \caption{Residuals of the best phase-connected timing solution including the \cha\ measured proper motion, fixing the position to the astrometric solution (upper panel) and allowing it to vary (lower panel).
 \label{fig:timing}
 }
\end{figure*}

A phase-coherent timing solution was then obtained by iteratively fitting the phases of the ToAs with the expression

\begin{equation}
\phi(t) = \phi_0 + \nu (t-t_0) + \frac{1}{2} \dot\nu (t-t_0)^2 + \frac{1}{6} \ddot\nu (t-t_0)^3,
\label{eq:phase}
\end{equation}
 \noindent
where $\nu$, $\nudot$, $\nuddot$ are the spin frequency and its time derivatives, and $t_0$ is
the epoch listed in Table~\ref{tab:timing}.
We started with the ToAs nearest to $t_0$, using only the linear term, and progressively included the other groups, and added higher order terms when observed phases began to deviate significantly from the prediction.
We obtained an acceptable fit ($\chi^2$/DoF = 321.1/219 = 1.47) with the parameters listed in   Table~\ref{tab:timing}.

In another analysis, we allowed the position
to be a free parameter while retaining the Chandra measured proper motion.  For this purpose, we calculated the 223 ToAs in the spacecraft-topocentric frame instead of the solar system barycentric frame, and we fitted them with the PINT software \citep[see also \citealt{2019ApJ...874..160D,ho22}]{2019ascl.soft02007L}. We use on-board GPS measurements, recorded every 10~s, to interpolate the position and velocity of NICER with respect to the geocenter at each ToA.

This results in a slightly
better fit ($\chi^2$/DoF = 285.7/217 = 1.32), with the residuals shown in the bottom panel of Figure~\ref{fig:timing}.  All timing parameters remain the same within errors, except that the fitted position at MJD 60374, R.A.=$14^{\rm h}12^{\rm m}56^{\rm s}\!\! .18(2),$ Decl.=$+79^{\circ}22^{\prime}03^{\prime\prime}\!\! .33(4)$, differs by $0^{\prime\prime}\!\! .25$ from the astrometric position.   In both fits, timing noise is evident as systematic residuals at the $1-2$~ms level, an effect that can bias a timing-derived position.  The  difference between the astrometric and timing positions, while statistically significant, is consistent with being due to timing noise.

While the X-ray timing also can constrain the proper motion, the current timing accuracy is insufficient for a significant measurement. Simulations using PINT indicate that at the current timing accuracy, the timing baseline would have to be extended by at least 3 decades before the precision of a proper motion constraint from timing would exceed that of the current astrometric measurement.

The second frequency derivative itself is evidently dominated by timing noise, as it is for most pulsars; the braking index $n=\brake=-505$ is much greater in absolute value than the dipole spin-down value of 3.  As discussed in \citet{mer21}, the timing noise of Calvera is higher than average, but still consistent with the distribution of values seen in normal radio pulsars with similar $\pdot$.

Because of the proper motion, the observed spin period derivative is affected by the Shklovskii effect \citep{1970SvA....13..562S}. However, this is 3 orders of magnitude smaller than the observed $\pdot$, so it is negligible.\\[10px]

\begin{table}
\centering \caption{Timing parameters  
\label{tab:timing}}
\begin{tabular}{lc}
\hline
\hline
Epoch of position (MJD) & 60374 \\
R.A. (J2000.0)\tablenotemark{a} & $14^{\rm h}12^{\rm m}56^{\rm s}\!\! .126(8)$  \\
Decl. (J2000.0)\tablenotemark{a} &  $+79^{\circ}22^{\prime}03^{\prime\prime}\!\! .54(2)$  \\
R.A. proper motion\tablenotemark{a}, $\mu_{\alpha}\cos\delta$ & $78.1\pm2.9$ mas\,yr$^{-1}$ \\
Decl. proper motion\tablenotemark{a}, $\mu_{\delta}$  &  $8.0\pm3.0$ mas\,yr$^{-1}$ \\
Epoch of timing, $t_0$ (MJD TDB) & $59240.0$ \\
Span of timing (MJD) & $58011-60247$  \\
Frequency, $\nu$      &  $16.89208599648(6)$ Hz \\
Frequency derivative, $\nudot$ & $-9.412042(11)\times10^{-13}$ Hz s$^{-1}$              \\
Frequency second derivative, $\nuddot$ & $-2.650(7)\times10^{-23}$ Hz s$^{-2}$  \\
Root-mean-square of residuals & $1.455$ ms \\
Number of ToAs & $223$ \\
$\chi^2$/DoF    &   $321.1/219$     \\[3pt]
\hline
\end{tabular}
\raggedright
\tablenotetext{a}{Chandra astrometric parameter held fixed.}
\end{table}

\section{Discussion} \label{sec:discussion}

\begin{figure}
\centering
\includegraphics[width=1\columnwidth]{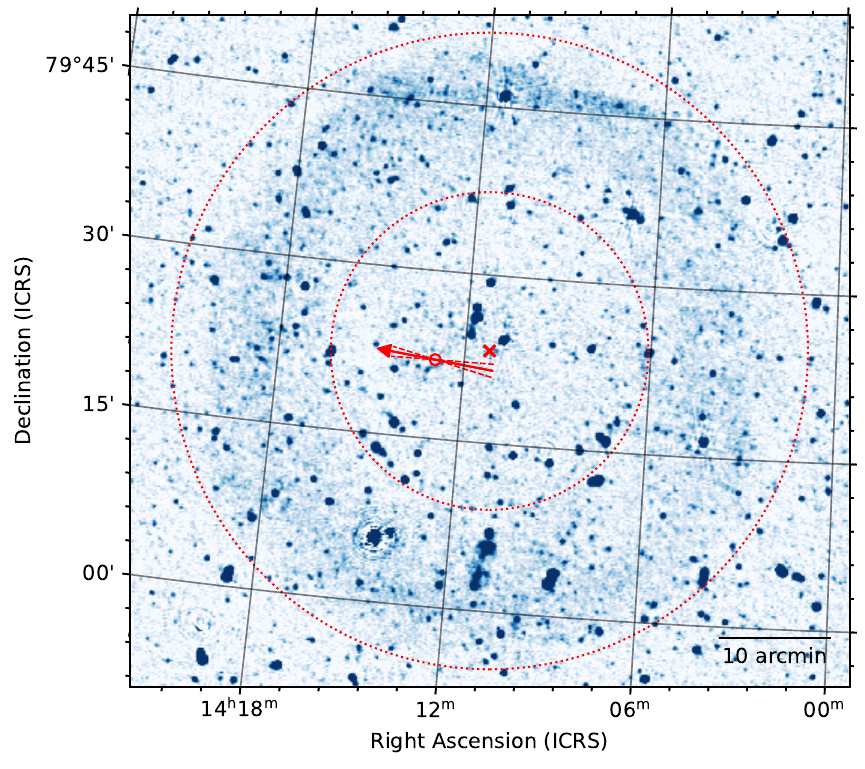}
 \caption{
LOFAR image of SNR G118.4+37.0 at $120-168$ MHz.
The cross marks the geometrical center of the SNR. The arrow indicates the proper motion of Calvera (currently at the position marked by the small circle), with dashed lines indicating the 3$\sigma$ uncertainty on position angle. The length of the arrow correspond to the distance covered by Calvera in $\pm4$ kyr. The large dotted circles, which delimit the region of diffuse radio emission, have  radii of $14^{\prime}$ and $29^{\prime}$.
\label{fig:snr}}
\end{figure}

The comparison of two \cha\ HRC-I data sets obtained in 2014 and 2024 has allowed us to detect the proper motion of Calvera with high statistical significance. The pulsar is moving at 
$\mu=78.5 \pm 2.9$ mas\,yr$^{-1}$ 
in position angle $84^{\circ}\!\! .1 \pm 2^{\circ}\!\! .2$. 
The proper motion direction differs from that previously obtained from the 2007 and 2014 observations. We attribute this discrepancy to the relatively small statistical significance ($<3\sigma$) of the previous measurement ($\mu=69\pm26$ mas\,yr$^{-1}$ at P.A.$=216^{\circ}\pm23^{\circ}$, \citealt{hal15}).

The proper motion direction, away from the center of the radio ring (see Figure~\ref{fig:snr}) supports the hypothesis that the latter is the SNR associated with Calvera. The remarkably circular shape of the radio emission allows a good determination of its center at R.A.=$14^{\rm h}11^{\rm m}12^{\rm s}\!\! .6$, Decl.=$+79^{\circ}23^{\prime}15^{\prime\prime}$ \citep{2022A&A...667A..71A}.
If Calvera was born at this position, its proper motion implies an age of $\approx\!4$~kyr.  Although the $\pm1\sigma$ range of trajectories doesn't actually overlap this position, the geometric center, even of a circular SNR, is not necessarily the location of the explosion \citep{wil13}.  Even assuming a conservative $8^{\prime}$ displacement of the explosion from the geometric center, it is clear that Calvera's age is $<10$~kyr, much smaller than its characteristic age of $\tau_c=285$~kyr. This implies that the pulsar was born spinning close to its present period.

In principle, the association with a SNR can provide some clues about the distance of Calvera.
The empirical relation between age and diameter recently derived by \citet{2023ApJS..265...53R} yields a diameter of $17.5\pm2.5$ pc for a 4 kyr old SNR, hence a distance of $\sim\!1$~kpc. 
However, there is a large dispersion around this average relation (see Figure 5 of these authors), with diameters spanning $\sim\!5$ pc to $\sim\!80$ pc, for ages between $\sim\!4$ and $\sim\!10$ kyr. 
Therefore,  the distance inferred from the SNR angular diameter is poorly constrained and still compatible with the range $3.1-3.8$ kpc  derived by \citet{mer21} modeling the thermal X-ray emission of Calvera.

Contrary to the previous results, the new proper motion indicates that Calvera is moving toward lower Galactic latitude, with tangential velocity $v_t=(372 \pm 14)~d_{1\,{\rm kpc}}$ km\,s$^{-1}$.
Taking into account the peculiar motion of the Sun and Galactic disk rotation would  give a velocity of 370~km\,s$^{-1}$ with respect to the local standard of rest of the pulsar (with proportionally larger corrections at larger distance). 
Such a velocity is most likely a kick acquired at birth and carries no clear evidence of the motion of the progenitor star.  Also, the sign of Calvera's  velocity on the axis perpendicular to the Galactic plane, $v_z$, depends on its unknown radial velocity $v_r$ as well as the latitude component $\mu_b$ of its proper motion:

\begin{equation}
v_z=d\,\mu_b\,{\rm cos}\,b+v_r\,{\rm sin}\,b.
\end{equation}
For $d=1$~kpc, if $v_r>+237$~km~s$^{-1}$ then Calvera's $v_z$ is away from the plane.
As to the origin of the progenitor, a $\approx\!10\,M_{\sun}$ star with a main-sequence lifetime of $\approx\!3\times10^7$~yr could travel $1$~kpc at a velocity of $33$~km\,s$^{-1}$, typical of runaway B stars. 

If the actual distance to Calvera is $\gsim\!3$~kpc as derived by the thermal X-ray modeling, then $v
_t\gsim1116$~km\,s$^{-1}$.  This is only slightly larger than that of two high-velocity pulsars with  good measurements of both parallax and proper motion, namely  PSR B2224+65 in the Guitar Nebula ($v_t=765^{+158}_{-98}$ km\,s$^{-1}$, \citealt{del19}) and PSR B1508+55 ($v_t=963^{+61}_{-64}$ km\,s$^{-1}$, \citealt{cha09}). 
The large distance would also place more stringent constraints on the properties of the runaway progenitor star that could have reached its location.

\cite{hal13} considered Calvera a possible descendant of the so-called central compact objects (CCOs), those $\approx\!10$ neutron stars in SNRs that are detected only in {\it thermal} X-rays \citep{got13}. Now, with the compelling association of a SNR with Calvera, the pulsar is, in some sense, an actual CCO.  The three original CCOs in which pulsations have been observed (see Figure~\ref{fig:ppdot}) have dipole magnetic fields of only $10^{10-11}$~G and characteristic ages of $\sim\!10^8$~yr; therefore, they hardly move in the $P$-$\dot P$ diagram. The absence of evidence for a much larger population of aged CCOs could be explained by magnetic field growth \citep{ho11,ho15}, in which an initial dipole field that was promptly buried by supernova fallback material of $\sim\!10^{-4}\,M_{\sun}$ diffuses out on a timescale of $\sim\!10^5$~yr, causing the pulsar to move upward in the $P$-$\dot P$ diagram.
The relatively weak dipole magnetic field of Calvera among the rotation powered pulsars suggests that its growth may continue.  

While the ages of SNRs hosting CCOs span a large range from $\approx\!350$~yr of Cas A \citep{2001AJ....122..297T,2006ApJ...645..283F} to $\sim\!27$~kyr of HESS J1731$-$347 \citep{2008ApJ...679L..85T}, the hosts of the three pulsating CCOs fall in the narrower range of $4-7$~kyr \citep{got13}.  These are comparable to the age of Calvera newly inferred here from its proper motion.  Under the CCO field-growth hypothesis for Calvera, this could imply that its diffusion time is shorter than that of the other three CCOs, probably because the fallback mass was smaller.

\section{Conclusions} \label{sec:conclusions}

Thanks to a new Chandra observation specifically planned to derive a precise position of the Calvera pulsar, we measured its proper motion with high  statistical significance. The angular velocity of $78.5\pm2.9$ mas\,yr$^{-1}$ at position angle of   $84^{\circ}\!\! .1 \pm 2^{\circ}\!\! .2$,  gives strong support to the association of Calvera with the recently discovered supernova remnant G118.4+37.0. 

We can thus consider Calvera as a new member of the CCO class. Its  dipole magnetic field  is a factor $\sim\!5$ to $15$ higher than that of the three other members of this class with measured  $P$ and $\pdot$, which suggests a larger variety in the birth properties and early evolution of CCOs.

The high Galactic latitude of Calvera is consistent with a B type runaway progenitor, but the pulsar proper motion, likely resulting from a SN kick, bears no information on the origin of the progenitor star. Further multi-wavelength observations of the now securely associated SNR can help to better constrain the distance of Calvera and to assess if its  velocity is in the high end of the pulsar velocity distribution, as implied from the $\sim\!3$ kpc distance estimated from its thermal X-ray emission.

\acknowledgments 
MR and SM acknowledge INAF support through the Large Grant ``Magnetars'' (P.I. S. Mereghetti) of the ``Bando per il Finanziamento della Ricerca Fondamentale 2022''.
Support for this work was provided to JPH and EVG through Chandra Award SAO GO3-2405X issued by the Chandra X-ray Observatory Center, which is operated by the Smithsonian Astrophysical Observatory for and on behalf of NASA under contract NAS8-03060.

\bibliographystyle{aasjournal} 
\bibliography{biblio}

\end{document}